\documentclass[11pt,oneside]{article}
\usepackage{enumerate}
\usepackage{amsmath}
\usepackage{amssymb}
\usepackage{geometry}               
\usepackage{pdfpages}          
\usepackage{graphicx}
\usepackage{amssymb}
\usepackage{float}
\usepackage{hyperref}
\usepackage{xcolor}
\usepackage{enumitem}
\usepackage{tikz}
\usetikzlibrary{matrix}

\newcommand{\KW}{\medskip\noindent{\bf Keywords: }}
\newcommand{\AMS}{\medskip\noindent{\bf 2010 Mathematics Subject Classification: }}

\title{A New Cryptographic Approach:\\
 Iterated Random Encryption (IRE)}
\author{Osvaldo Skliar\thanks{Escuela de Informática, Universidad Nacional, Heredia, Costa Rica. E-mail: oskliar@costarricense.cr. ORCID: 0000-0002-8321-3858.}
\and Sherry Gapper\thanks{Universidad Nacional, Heredia, Costa Rica. E-mail: sherry.gapper.morrow@una.ac.cr. ORCID: 0000-0003-4920-6977.} \and 
Ricardo E. Monge\thanks{Universidad de Costa Rica, San José, Costa Rica. E-mail: mongegapper@gmail.com. ORCID: 0000-0002-4321-5410.}
}
\date{\today}

\begin{document}

\maketitle

\begin{abstract}
\noindent A new cryptographic approach -- Iterated Random Encryption (IRE) -- is presented here. Although it is very simple, and easy to implement, it provides a very high level of security. According to this approach, a sequence of operations applied to a message ($M$) yields the encrypted message ($M_E$). In that series of operations, the one with the most important role is operation 6, which involves a random binary sequence (RBS) generated by using the Hybrid Random Number Generator (HRNG) or the Mathematical Random Number Generator (MRNG). A sequence of anti-operations applied to $M_E$ makes it possible to recover $M$.

\end{abstract}
\KW cryptography, random number generators, random binary sequence

\AMS 11T71, 68P25, 94A60\\

\section{Introduction}
The objective of this article is to present a new cryptographic approach. That proposed here -- Iterated Random Encryption (IRE) -- is a novel approach. It is  different from those currently predominant, such as RSA (Ron \textit{R}ivest, Adi \textit{S}hamir, Leonard \textit{A}dleman), which are based on the tremendous practical difficulty of factoring the product of two very large prime numbers.

Introductory information about cryptography may be found, for example, in \cite{cr1}, \cite{cr2}, \cite{cr3} and \cite{cr4}.

The terminology used is expected to make this article clear even for readers less familiar with cryptography and cryptanalysis. An effort has been made for basic notions of mathematics (such as permutation) and of computer science (such as bit, byte and pixel) to suffice for a clear understanding of the essential ideas in the article. In the title of this paper, mention is made of an ``approach", rather than of a ``method" because the main ideas explained here may be implemented in various ways.

The effective use of this approach leads to the generation of certain products. Each product is composed essentially of a) a computer program oriented toward encrypting information; and b) a computer program which makes it possible to decrypt the information encrypted by the program referred to in a) above. Those products will be referred to as $S_1, S_2, S_3, \ldots$ ($S$ for Software).

Each $S_i$, for $i=1,2,3,\ldots$, can have various users. It is accepted that there will be a minimum of 2 users for each $S_i$, and either of them can encrypt information ($M$), which may consist of either a text or an image ($M_E$), and send the corresponding encrypted message to the other user. The latter will be able to decrypt the message using the computer program mentioned above in b).

 $C_i$ stands for the set of users of $S_i$. Only the users who are members of a particular set of users $C_i$, for $i=1,2,3\ldots$, will be able to use $S_i$. Each user can be a member of more than one $C_i$; for instance, of $C_3$ and $C_7$. Obviously, as a member of $C_3$, a person will use $S_3$, and as a member of $C_7$, that person will use $S_7$.\\

\section{How to Develop the Software Described}
An explanation of how to develop a given $S_i$, for $i=1,2,3\ldots$, will be provided in this section. As seen above, $M$ stands for the information to be encrypted, and $M_E$ for the encrypted message. To encrypt a message, one must follow a sequence of operations.

\subsection{Operation 1}
It is admitted that the information to be encrypted $M$ is expressed by a sequence of bytes. Each byte is composed of a sequence of 8 bits. Each bit can be represented by a zero (0) or by a one (1).

Given that each byte is composed of a sequence of 8 bits, the number of different bytes is $2^8$ (that is, 256).

The number of bytes composing $M$ is then counted. If that number of bytes is a natural number $N_B$ such that $N_B$ is less than 10 (that is, $N_B<10$), a number of bytes equal to $(10-N_B)$, such that each of the latter symbolizes the whitespace character, is added to that sequence of bytes. Thus, suppose that, for example, $M$ is comprised of a sequence of 8 bytes:

\begin{equation*}
M:\Big[\parbox{2em}{\centering first byte}\Big]\Big[\parbox{3em}{\centering second byte}\Big]\Big[\parbox{3em}{\centering third byte}\Big] \cdots \Big[\parbox{4em}{\centering seventh byte}\Big]\Big[\parbox{3em}{\centering eighth byte}\Big]  
\end{equation*}\

To that sequence $M$, 2 bytes ($10-8=2$) are added. Within the symbolic system used, each of them symbolizes the whitespace character. In this way, the sequence $M_1$ of bytes is obtained:

\begin{equation*}
M_1: \Big[\parbox{2em}{\centering first byte}\Big] \Big[\parbox{3em}{\centering second byte}\Big] \cdots \Big[\parbox{3em}{\centering eighth byte}\Big]  \Big[\parbox{3em}{\centering ninth byte}\Big]   \Big[\parbox{3em}{\centering tenth byte}\Big]  
\end{equation*}\

In $M_1$, the ``ninth byte" in the sequence is equal to the ``tenth byte", and within the symbolic system used, both represent the whitespace character.

If that number of bytes in the sequence comprising $M$ is equal to 10 or greater than 10 (that is, $N_B\geq 10$), the sequence $M_1$ of bytes is considered equal to $M$.

It can be accepted that the operation carried out on $M$ (called ``operation 1") is applied by an operator called $O_1$. The action of $O_1$ on $M$ is symbolized as follows:
\begin{equation}
O_1\{M\}=M_1
\end{equation}

For operation 1, as for the following operations to be considered here, there is a corresponding ``anti-operation". In this case, "anti-operator 1" is called $AO_1$. The action of $AO_1$ on $M_1$ is symbolized as follows:
\begin{equation}
\tag{$1'$}
AO_1\{M_1\}=M
\end{equation}

If it is detected, for example, that $M_1$ is composed of a sequence of 10 bytes and that each of the last two bytes in that sequence symbolizes the whitespace character, then $AO_1$, when acting on $M_1$, will eliminate the last two bytes in that sequence, and $M$ will be recovered.

If, on the other hand, $M_1$ is composed of a number of bytes greater than 10, the operator $AO_1$, when acting on $M_1$, will leave $M_1$ unchanged. In that case, $M_1$ will be the same as $M$.

\subsection{Operation 2}

Suppose that the 256 bytes mentioned above are in a sequence of binary numbers in ascending order. That ordered sequence of 256 bytes will be known as ``$s$".

Operation 2 consists of a) randomly carrying out a permutation of the 256 bytes in that ordered sequence, and b) establishing a one-to-one correspondence between each byte of $s$ and the byte which corresponds to it according to the permutation carried out.

The operator that acts on $s$ to generate a new sequence of bytes will be known as ``random permutation operator" ($O_{R.P.}$).
\begin{equation}
O_{R.P.}\{s\}=s_p
\end{equation}

Thus, $s_p$ is the ``new sequence" of bytes generated, such that the first byte of $s_p$ corresponds to the first byte of $s$, the second byte byte of $s_p$ corresponds to the second byte of $s$, the third byte of $s_p$ corresponds to the third byte of $s$, and so on successively.

In this case, the corresponding ``anti-operator" acts on $s_p$ and generates $s$.
\begin{equation}
\tag{$2'$}
AO_{R.P.}\{s_p\}=s
\end{equation}

Hence, for example, if when carrying out this permutation of the bytes in $s$, bytes (\texttt{01101100}) and (\texttt{01010101}) correspond to bytes (\texttt{01001101}) and (\texttt{01110001}) respectively, then when carrying out the inverse permutation, symbolized by (2'), bytes (\texttt{01001101}) and (\texttt{01110001}) would correspond to (\texttt{01101100}) and (\texttt{01010101}) respectively.

There are $256!$ possible permutations of the 256 bytes in $s$. Some of these permutations will assign to some byte in $s$ (or to more than one byte in $s$) the same byte (or bytes) in $s_p$. In particular, one of these $256!$ permutations is that which assigns to each byte in $s$ the same byte of $s_p$. (In this case $s=s_p$.) The probability that the randomly generated permutation would be this specific one is very low: $\tfrac{1}{256!}$.

\subsection{Operation 3}

Consider the sequence $M_1$ of bytes mentioned in section 2.1. In ``operation 3", each byte of $M_1$ is replaced by the corresponding byte according to the operation symbolized in (2):
\begin{equation}
O_3\{M_1\}=M_3
\end{equation}\
The corresponding anti-operation, whose nature appears obvious, can be symbolized as:
\begin{equation}
\tag{$3'$}
AO_3\{M_3\}=M_1
\end{equation}
\subsection{Operation 4}

Therefore, $M_3$ is a sequence of at least 10 bytes.

First, a description will be provided of how to proceed if $M_3$ is composed of only 10 bytes; and then of how to proceed if $M_3$ is a sequence of more than 10 bytes.

In the first case, the 10 bytes comprising $M_3$ are numbered according to the order they are in, and not according to their numerical values in binary arithmetic. Hence, the following sequence of bytes is obtained:\\

\noindent$s_1$: byte 1, byte 2, byte 3, byte 4, byte 5, byte 6, byte 7, byte 8, byte 9, byte 10\\

Next, one of the randomly selected $10!$ possible permutations of those 10 bytes of $s_1$ is carried out. Thus, in $s_2$ the new order of those 10 bytes in $s_1$ could be, for example, as follows:\\

\noindent$s_2$: byte 8, byte 3, byte 7, byte 4, byte 1, byte 10, byte 2, byte 9, byte 6 byte 5\\

Suppose now that $M_3$ is a sequence of more than 10 bytes; for example, of 15 bytes:\\

\noindent$s_3$: byte 1, byte 2, byte 3, byte 4, byte 5, byte 6, byte 7, byte 8, byte 9, byte 10, byte 11, byte 12, byte 13, byte 14, byte 15\\

Let us admit that there is a ``window'' which makes it possible to visualize exactly 10 consecutive bytes; and that this ``window" may be shifted, and that when shifted, it does not take with it the bytes which that window has made it possible to visualize.

Suppose that initially this shiftable ``window" makes it possible to visualize the first 10 bytes of $M_3$. (Note that this sequence $M_3$ of 15 bytes was also called $s_3$.)

The first 10 bytes are subjected to an arbitrarily selected permutation. To simplify this description, let us accept that this random permutation is the same as that which went from sequence $s_1$ to sequence $s_2$. (It could have been any other of the $10!$ possible permutations.) Once that random permutation is complete, that ``window" is shifted one byte. As a result of that shift, the new sequence of bytes which will be found in the ``window" is as follows:\\

\noindent$s_4$: byte 3, byte 7, byte 4, byte 1, byte 10, byte 2, byte 9, byte 6 byte 5, byte 11\\

Then the 10 bytes of $s_4$ are renumbered: the first byte of $s_4$ (byte 3) is again called ``byte 1"; the second byte of $s_4$ (byte 7) is called ``byte 2"; the third byte of $s_4$ (byte 8) is called ``byte 3", and so forth, until the tenth byte of $s_4$ (byte 11) is called ``byte 10". Once the renumbering is completed, the ``new window" appears as follows:\\

\noindent$s_5$: byte 1, byte 2, byte 3, byte 4, byte 5, byte 6, byte 7, byte 8, byte 9, byte 10\\

The preceding sequence $s_5$ of bytes is then subjected to the same permutation carried out with the first 10 bytes of $M_3$. This operation will result in a sequence of bytes that will be again numbered as follows:\\

\noindent$s_6$: byte 8, byte 3, byte 7, byte 4, byte 1, byte 10, byte 2, byte 9, byte 6 byte 5\\

The iteration process continues: 1) a new shift of the ``window" by one byte, in the increasing order of the bytes of $M_3$; 2) the renumbering of the bytes that can be visualized in the new position of the window; and 3) the reiteration of the same permutation to which the first 10 bytes of $M_3$ were subjected.

In the case considered of a sequence $M_3$ of 15 bytes, the window was shifted one byte 5 times. There were 5 processes of renumbering bytes and 6 permutations, the first of which is that applied to the first 10 bytes of $M_3$.

In general, if a sequence $M_3$ of $N_B$ bytes is considered, then there would be ($N_B-10$) shifts of the ``window", of one byte each; ($N_B-10$) processes of renumbering bytes, and ($N_B-10+1$) permutations of the type described of 10 bytes of each of the successive windows that must be taken into account.

Operation 4 may be symbolized as follows:
\begin{equation}
O_4\{M_3\}=M_4
\end{equation}

It is easy to observe that the inverse process, which goes from $M_4$ to $M_3$, can be symbolized as:
\begin{equation}
\tag{$4'$}
AO_4\{M_4\}=M_3
\end{equation}

\subsection{Operation 5}

According to the above, $M_4$ is a sequence of at least 10 bytes: $N_B\geq 10$. $M_4$. Given that each byte is a sequence of 8 bits, it is also possible to consider that $M_4$ is composed of a sequence of at least 80 bits: $N_b\geq 80$.

With that sequence of bits a procedure like that specified in section 2.4 is used, with the following differences: 1) The random permutation considered will not be of 10 bytes but rather of 80 bits; and 2) the shifts of an 80-bit  ``window" will not be shifted one \textit{byte} but rather one \textit{bit}, each time it occurs.

It is clear that for $M_4$, the following equation will be valid: $N_b=8N_B$.

In operation 5, ($N_b-80$) shifts of one bit will be carried out. In addition, ($N_b-80$) processes of renumbering bits will take place. Note that if $N_b=80$, the number of shifts will be equal to 0; that is, no shift will take place.

The same random permutation of 80 bits, which may be one of any of $80!$ possible permutations of those bits, will be applied to each of the successive ``windows" of 80 bits. The number of these permutations will be equal to ($N_b-80+1$). Note that if $N_b=80$ (that is, if $M_4$ is a sequence of only 80 bits), only one permutation will be carried out: that of 80 bits.

Operation 5 may be symbolized as:
\begin{equation}
O_5\{M_4\}=M_5
\end{equation}

Hence, it is thus easy to see that the inverse process, which goes from $M_5$ to $M_4$, can be symbolized as:
\begin{equation}
\tag{$5'$}
AO_5\{M_5\}=M_4
\end{equation}

\subsection{Operation 6}
In operation 6, two random number generators play an essential role (or at least one of them does): the Hybrid Random Number Generator (HRNG) \cite{hrng}, and the Mathematical Random Number Generator (MRNG) \cite{mrng}.

Regarding the HRNG, in \cite{sghrn} the following has been stated: ``This HRNG constructs indisputable random numbers which may be used in the field of cryptography and many other computer applications starving for a random seed" (p. 312).

The article on the MRNG was mentioned in the MIT Technology Review \cite{mit}.

A random binary sequence (RBS) can be generated by using the HRNG or the MRNG. This type of sequence used to generate the results presented in section 3 was composed of ($30{,}064 \times 10^6$) bits. By using these random number generators, there are no technical difficulties to obtain binary sequences composed of much larger numbers of bits.

Before specifying how the binary sequence is implemented effectively, the main idea of how it is used will be provided below. 

Let there be a fragment of the sequence $M_5$ of bits, such as the following:\\

\noindent$s_7$: 1,0,1,1,0,0,1,0,1,0\\

It can be seen that the preceding fragment of $M_5$ is a sequence of 10 bits. 

To that fragment of $M_5$, another fragment (also composed of 10 bits from the RBS), is assigned, having been generated by using the HRNG or the MRNG.

Suppose that the fragment, also composed of 10 bits, from the RBS, generated by using the HRNG or the MRNG is as follows:\\

\noindent$s_8$: 1,0,0,1,1,0,0,0,0,1\\

Figure 1 displays how a one-to-one correspondence is established between each of the bits comprising the fragment of $M_5$ and each corresponding bit of the above fragment of RBS.\\

\begin{figure}[H]
  \centering
  \begin{tikzpicture}
      \matrix (m) [matrix of  nodes] {
      Fragment of sequence $s_7$ of $M_5$:$\quad$ & 1 & 0 & 1 & 1 & 0 & 0 & 1 & 0 & 1 & 0 &  \\
       Fragment of sequence $s_8$ of RBS:$\quad$ & 1 & 0 &0 & 1 & 1 & 0 & 0 & 0 & 0 & 1	 &  \\
      }; 
      \draw [dashed] (m-1-2.north west) -- (m-2-2.south west);
      \draw [dashed] (m-1-3.north west) -- (m-2-3.south west);
      \draw [dashed] (m-1-4.north west) -- (m-2-4.south west);
      \draw [dashed] (m-1-5.north west) -- (m-2-5.south west);
      \draw [dashed] (m-1-6.north west) -- (m-2-6.south west);
      \draw [dashed] (m-1-7.north west) -- (m-2-7.south west);
      \draw [dashed] (m-1-8.north west) -- (m-2-8.south west);
      \draw [dashed] (m-1-9.north west) -- (m-2-9.south west);
      \draw [dashed] (m-1-10.north west) -- (m-2-10.south west);
      \draw [dashed] (m-1-11.north west) -- (m-2-11.south west);
      \draw [dashed] (m-1-11.north east) -- (m-2-11.south east);
\end{tikzpicture} 
\caption{One-to-one correspondence between bits}
\end{figure}\

As seen in figure 1, the first bit of the fragment of the RBS (i.e., 1) corresponds to the first bit of the fragment of $M_5$ (i.e., 1); the second bit of the fragment of the RBS (i.e., 0) corresponds to the second bit of the fragment of $M_5$ (i.e., 0); the third bit of the fragment of the RBS (i.e., 0) corresponds to the third bit of the fragment of $M_5$ (i.e., 1); and so on successively until the tenth bit of the fragment of the RBS  (i.e., 1) corresponds to the tenth bit of the fragment of the $M_5$ (i.e., 0).

Attention is given below to all of the pairs of bits composed of one of the bits of the fragment of $M_5$ and its corresponding bit in the fragment of RBS. These pairs of bits are as follows:
\begin{equation*}
(1,1) \: (0,0) \: (1,0) \: (1,1)\: (0,1) \: (0,0) \: (1,0) \: (0,0) \: (1,0) \: (0,1)         
\end{equation*}

It is accepted that each of these pairs of bits generates a bit according to one of the following rules:
\begin{enumerate}[label=\alph*)]
\item If both bits in a pair of bits are equal, a 1 (one) is generated; and if they are not equal, a 0 (zero) is generated; or
\item If both bits in a pair of bits are equal, a 0 (zero) is generated; and if they are not equal, a 1 (one) is generated.
\end{enumerate}

Suppose that b, the second of the above rules, has been chosen. The result obtained when opting for b is displayed in figure 2.\\

\begin{figure}[H] 
  \centering
\begin{tikzpicture}\

      \matrix (m) [matrix of  nodes] {
      Fragment of sequence $s_7$ of $M_5$:$\quad$ & 1 & 0 & 1 & 1 & 0 & 0 & 1 & 0 & 1 & 0 & \\
      Fragment of sequence $s_8$ of RBS:$\quad$ & 1 & 0 &0 & 1 & 1 & 0 & 0 & 0 & 0 & 1	 &  \\
      Fragment of sequence $s_9$ of result:$\quad$ & 0 & 0 & 1 & 0 & 1 & 0 & 1 & 0 & 1 & 1	 &  \\
      }; 
      \draw [dashed] (m-1-2.north west) -- (m-3-2.south west);
      \draw [dashed] (m-1-3.north west) -- (m-3-3.south west);
      \draw [dashed] (m-1-4.north west) -- (m-3-4.south west);
      \draw [dashed] (m-1-5.north west) -- (m-3-5.south west);
      \draw [dashed] (m-1-6.north west) -- (m-3-6.south west);
      \draw [dashed] (m-1-7.north west) -- (m-3-7.south west);
      \draw [dashed] (m-1-8.north west) -- (m-3-8.south west);
      \draw [dashed] (m-1-9.north west) -- (m-3-9.south west);
      \draw [dashed] (m-1-10.north west) -- (m-3-10.south west);
      \draw [dashed] (m-1-11.north west) -- (m-3-11.south west);
      \draw [dashed] (m-1-11.north east) -- (m-3-11.south east);
      
\end{tikzpicture}
\caption{The result of applying rule b above}
\end{figure}\

It is evident that the process specified above must be applied to the entire sequence $M_5$ of bits, and not to only one fragment of it. This operation can be symbolized as follows:
\begin{equation}
O_6\{M_5\}=M_6=M_E
\end{equation}

As with the other operations above, it is clear that there is a sixth anti-operation, which when acting on the sequence of bits in $M_6$, generates the sequence $M_5$ of bits:
\begin{equation}
\tag{$6'$}
AO_6\{M_6\}=M_5
\end{equation}

Additional information will be provided below on the effective use of the encryption and decryption programs described for any RBS that may be generated with the HRNG or the MRNG, and used in operation 6.

First, the bits making up the RBS to be used in successive ``sixth operations" should be numbered. Thus, the first bit of the sequence is called bit 1, the second bit, bit 2, the third, bit 3, and so on successively until the $n^{th}$ bit of the sequence is called bit $n$. For example, the last bit of the RBS used to obtain the results presented in section 3 is called ($30{,}064\times 10^6$).

Second, each sequence of this type (that is, each RBS) is converted into a \textit{loop}. Hence, just as bit 36 is followed by bit 37, and bit 350,258 is followed by bit 350,259, an operation is carried out so that the last bit in the sequence is followed by the first bit. In the case of the software used to obtain the results presented in section 3, an operation is carried out so that bit $30{,}064\times 10^6$ is followed by bit 1.

By using a loop (or closed sequence) of bits, the bits of the RBS found in operation 6 are never ``used up".

Although it is unlikely that exactly the same fragment of the loop may be found in two different instances of operation 6, for security reasons that loop should be composed of a number of bits as large as the space available in the memory of the computing device will permit.

Suppose that in a given operation 6, the encryption program begins to use the RBS as of bit 6,723 inclusively. So that the encrypted message to be sent can be decrypted by another user of the same $S_i$, for $i=1, 2, 3, \dots$, something must be added; for example, that message must be preceded by certain information: the number of the initial bit of the fragment of the RBS used in operation 6. In this case, the number is 6,723. (Of course, in other cases the numbers added to the encrypted messages will usually be different from 6,723.)\\

\section{Results Obtained by Using the Encryption and Decryption Programs Described}

\subsection{Encryption and Decryption of a Text}

Consider the text in figure 3. This extract was taken from page 2 of one of our papers on random number generation \cite{mrng}. This page contains the initial section of the contents of the paper (the Introduction and the first part of Section 1).
\begin{figure}[H]
  \centering
\frame{\includegraphics[width=4.7in]{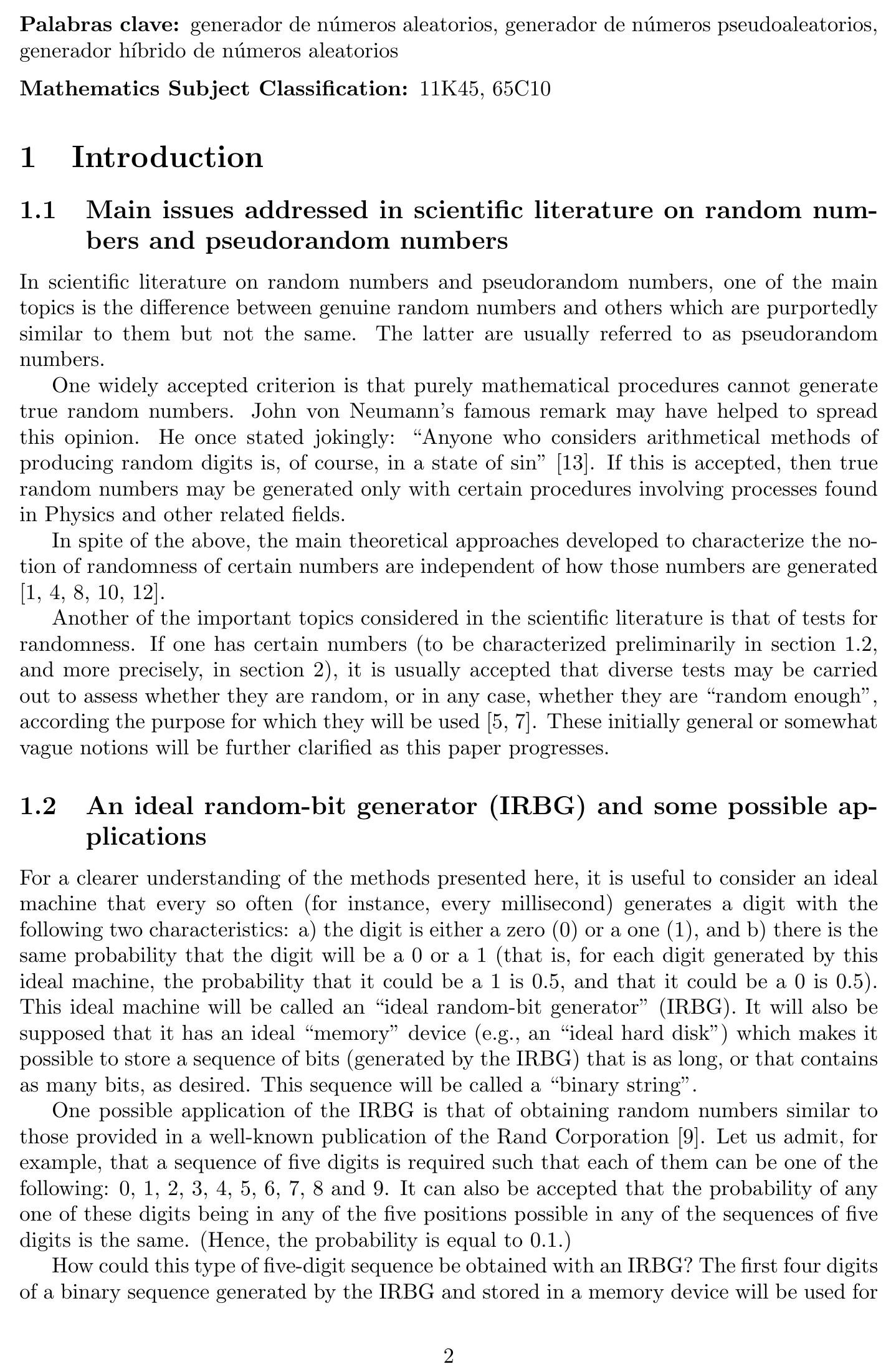}}
\caption{Text to be encrypted}
 \end{figure}
Once we use the IRE to process the "PDF binary sequence" corresponding to the text displayed in figure 3, we obtain a sequence of bits (grouped in bytes), the first 2,400 of which are reproduced below:\\

\begin{quote}
10100001 00001010 11110000 00000000 00001110 01101110 00111010 10011100 
01111111 00010110 00111010 10111100 11110011 10100111 10100101 01101011 
11001111 10110001 01100101 00101100 00011010 10010110 01010001 00100000 
01001110 11110101 01001000 10010010 00000101 10100110 01110101 11100000 
10000010 10101011 00010000 01000000 10000110 00001111 00111011 10001001 
01001110 00111000 10100001 01110000 00011110 00001011 00011110 10011011 
11001110 11101101 10011111 01010101 00100010 01101110 00110100 01100110 
10000111 10000111 00101010 00100111 10110011 01001110 00111010 10110111 
10001101 01011101 00011011 00101000 11110111 00111110 01100111 01100011 
10100101 10000101 11000100 11101100 10101100 10000100 00110110 00011101 
01001111 11111000 01011100 11101101 00111101 00100110 11111111 11110110 
10011011 10011011 10110000 00010011 00010111 10101100 01011011 00111110 
11101001 01101000 10010110 11110110 11101011 10100110 00010100 11000110 
00010001 10111001 10010101 01010011 10000000 00111111 00110001 01110000 
00000000 10101000 01011010 01110101 10110110 10001101 10101000 00110111 
10111110 01001100 11001011 11110010 10111101 10010000 11111110 00001110 
10111010 00100010 01010100 11111111 00011101 10111001 00111010 00101011 
11010010 11100000 11010111 01000111 10101101 00000101 01111111 00110001 
01010100 01000011 10111001 01100001 11001000 01000010 11100011 01100100 
10100100 01010011 00001011 11010111 00000000 01001111 01111100 10101011 
00111101 10110000 00000101 00010101 11001000 11101110 11110111 00100110 
01011100 11111001 01111111 00111010 01001111 11111000 01110001 11001100 
11010011 01110000 01100101 01000001 10000110 01010110 10001011 00110001 
01011101 00000101 11111011 11101001 11000100 10000111 11001001 01101110 
01101101 01111100 11010111 10110001 00111101 01011010 10110010 01110000 
00001110 11001111 10010101 11111111 00101100 01000100 01010111 01000010 
11100011 01011000 01000010 11101001 01110011 11110111 10000010 11011001 
01011011 00010100 11100110 11001000 10100001 11111001 00010110 11000001 
01101001 00100001 10001001 10101111 00011110 10001010 01101110 01111010 
10110111 01011010 11001100 11101010 01010111 11100001 01111000 10100111 
10110011 00100001 10101000 11000000 01111100 10011011 11010110 01111010 
11001000 01011101 01011111 00111111 00100101 11001001 00111010 00110100 
10010011 00111010 00011111 00100011 00000110 11110100 11110000 10101011 
01000010 00111100 11011100 11101000 10100110 11101101 11101101 10011000 
11000110 10000011 01001000 00100010 10110101 11110101 00011010 00000010 
10100000 00011001 00010000 00110110 01001100 01011000 00011101 01110111 
11000001 10011011 00111111 10010100 10010011 01000110 11010011 01110100 
00001000 00011111 10111010 11000100
\end{quote}

The use of the corresponding decryption program makes it possible to recover the original text provided above.

If that same text is encrypted again immediately or after other encryption processes, a different encrypted message is generated. The reason for that is the following: The initial bit of the fragment of the RBS used for that second occasion in operation 6 is different from the initial bit of the other fragment of the RBS used previously.

\subsection{Encryption and Decryption of an Image}

Consider the image composed of ($24\times 10^6$) pixels (in figure 4). It belongs to a gallery of digital art \cite{ec}.
\begin{figure}[H]
  \centering
  \includegraphics[height=4.7in]{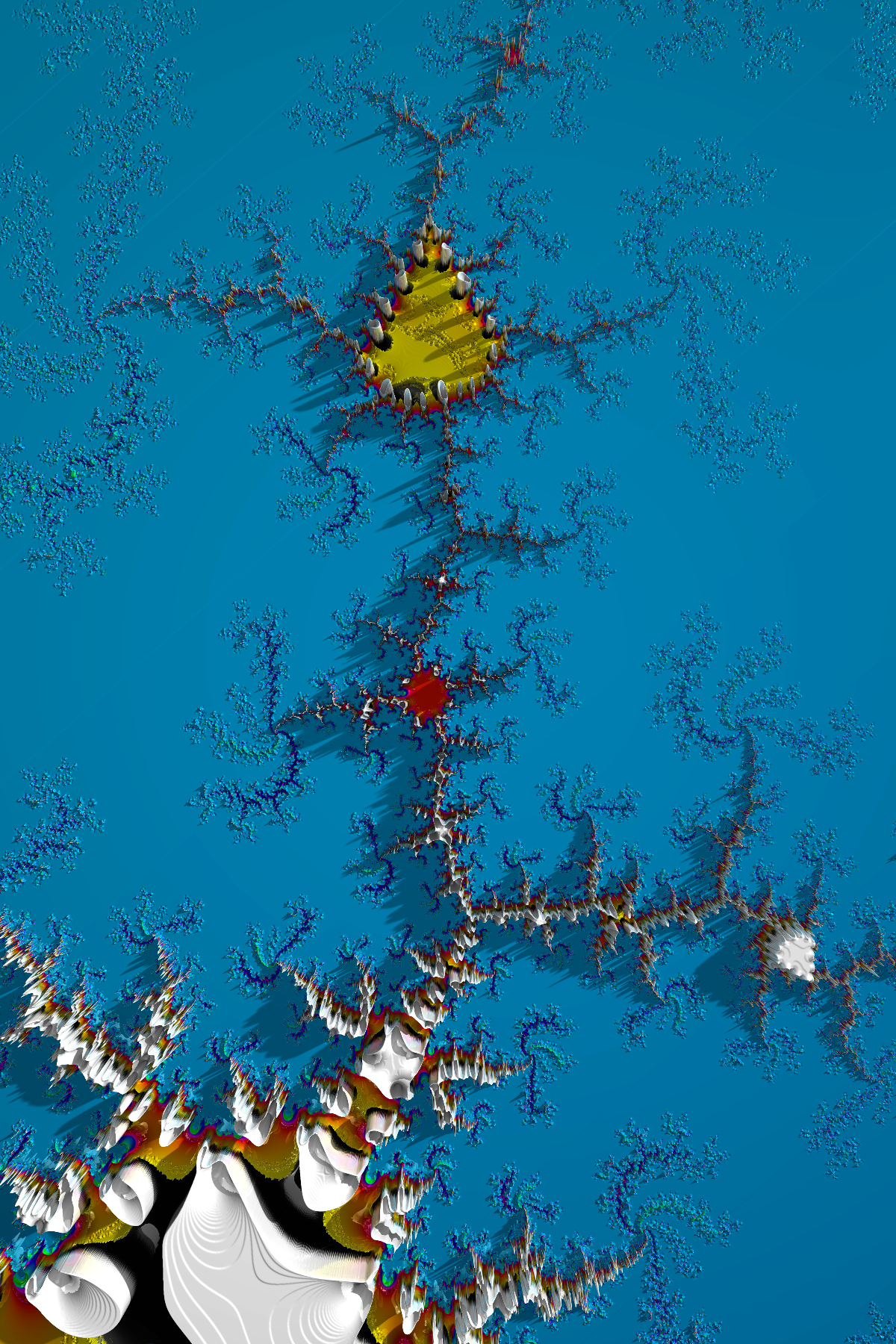}
\caption{Fractal image to be encrypted}
 \end{figure}
Each pixel bears a color which is the result of a combination of certain levels of intensity of three different colors: red (R), green (G), and blue (B).The levels of each of these colors can vary between 0 and 255. In other words, there are 256 levels of color intensity for each of the three colors. That makes it possible to select from $256^3$ (that is, 16,777,216) colors for each pixel in the image). The combination of the levels of 0 for all three colors yields a black pixel, and the combination of the levels of 255 for each color yields a white pixel.

The specification, for each pixel, of its corresponding levels of color requires 3 bytes: one for the level of intensity of color for each of the three colors mentioned above.

Given a) the number of pixels in the image considered ($24\times 10^6$), and b) the 3 bytes necessary to specify the information for each pixel, the image can be expressed or coded by a sequence of ($3\times 24\times 10^6$) -- that is, ($72\times 10^6$) -- bytes. That sequence can be encrypted using the approach described in section 2.

The decryption program, when operating on the encrypted message, makes it possible to recover the image in figure 4.

It would consume too much space to use a sequence of ($8\times 72\times 10^6$) --   that is, ($576\times10^6$) bits -- corresponding to the encrypted message. To give an idea below of the appearance of the encrypted message, only the first 800 bits are shown, grouped in bytes.

\begin{quote}
11101000 11000110 10101100 01101111 00010010 00110101 10001100 00101110 
11011000 10010111 11001101 00101001 11010011 11111110 01111111 00111100 
01111000 10011001 00100010 00101001 00111101 10000000 10100101 11001000 
11000010 11000100 00100101 10101111 11110111 01100111 00000111 11111111 
00010100 11101000 00000110 10111100 11000001 00111111 11010110 11111101 
11011111 11111001 10001011 11111000 10011101 00100001 11111010 01110000 
00111111 10000001 11001000 11001101 01100011 00011101 11100000 01110000 
10000001 00011110 00101011 00101111 10011101 01111000 00111110 10111000 
00000101 10100101 11001011 11110100 11010010 01000000 01110111 11011101 
10001011 01000001 00110000 10111111 10111000 10001101 01100100 01000000 
00000001 11001011 01101111 00110001 11011101 00111001 11010100 01011101 
00110000 11100000 11100110 10011010 01010111 01111011 11000000 10101100 
00111010 11011111 11000111 00111011 
\end{quote}

The computer equipment used to run the encryption and decryption programs described here has a six-core Intel Core i7 5820K processor, of which one of the cores was used. Each of these programs was developed in two different programming languages: a ) Microsoft Visual C\# 2017 for the Windows version and b) GNU C++ (using the QT Toolkit) for the multi-platform version.

There was no significant difference between the encryption and decryption time for the text presented in 3.1 -- both were less than 1 ms (one millisecond); nor was there a significant difference between the encryption and decryption time for the image presented in 3.2: both were less than approximately 3 ms (three milliseconds). The trials carried out so far encrypting and decrypting information with some $S_i$, for $i=1, 2, 3, \ldots$, show that these times increase linearly with the number of bytes in $M$.\\

\section{Discussion and Prospects}

As seen above, $O_1$, $O_{RP}$, $O_3$, $O_4$, $O_5$ and $O_6$ do not add information about the nature of the operators that they represent. The exact meaning of each of them is specified in sections 2.1, 2.2, 2.3, 2.4, 2.5 and 2.6, respectively. Anyone who, for reasons of mathematical rigor, prefers to eliminate them may do so because only a clear understanding of what they refer to is required.

Nevertheless, they serve to summarize the sequence of operations acting on $M$ (the message to be encrypted) and $M_E$ (the encrypted message itself).
\begin{equation}
O_6\{ O_5\{ O_4\{ O_3\{  O_1\{ M  \} \}  \}\} \}=M_6=M_E
\end{equation}

Note that $O_{RP}$ does not appear explicitly in (7), but the operation it represents is indispensable for operation 3. 

The sequence of anti-operations which, when acting on $M_6=M_E$, makes it possible to obtain $M$ once again can be symbolized as follows:
\begin{equation}
\tag{$7'$}
AO_1\{ AO_3\{ AO_4\{ AO_5\{  AO_6\{ M_E  \} \}  \}\} \}=M
\end{equation}\
Note that $AO_{RP}$ does not appear explicitly in (7'), but the anti-operation it represents is indispensable for the first anti-operation.

There is no doubt that operations 1, 2, 3, 4 and 5 each contribute to the security of the desired result. However, the most cryptographically important operation is the sixth one, in which the RBS obtained from the HRNG or the MRNG is used.

The computing device containing the programs for encrypting and decrypting messages should not be connected to the internet.

If a member of any $S_i$, for $i=1, 2, 3, \ldots$, receives an encrypted message from another member of the same $S_i$, that member may do the following: First, that person must copy the message onto a device which is suitable for the task and download the message in the computational device used to encrypt and decrypt messages; then the message may be decrypted.

If someone desires to respond with an encrypted message, that person must 1) write the reply and encrypt it in the encryption-decryption device, and 2) copy the encrypted message onto a computer with access to the Internet, and send it to its respective destination.

Although the effectiveness and efficiency of this approach will be analyzed elsewhere, certain preliminary remarks about its effectiveness are included below.

The diverse random permutations used in IRE are a serious obstacle for any cryptanalyst interested in deciphering a message encrypted by using that approach.

As indicated above, there are 256! possible permutations involved in operation 2 (described in section 2.2). If, when carrying out that permutation, a genuine random number generator is used, all of the possible permutations are equiprobable: Each one has a probability equal to $\tfrac{1}{256!}$ of being selected.

Operation 4 (described in section 2.4) implies a permutation of 10 elements (in this case, 10 bytes). If a genuine random number generator is used to carry out this permutation, all of the possible permutations are equiprobable: Each one has a probability equal to $\tfrac{1}{10!}$ of being selected.

Operation 5 (described in section 2.5) requires a permutation of 80 elements (in this case, 80 bits). If a genuine random number generator is used to carry out this permutation, all of the possible permutations are equiprobable: Each one has a probability equal to $\tfrac{1}{80!}$. of being selected.

Therefore, the probability of discovering what those three permutations are, by mere chance, is the product of the three: ($\tfrac{1}{256!}\times\tfrac{1}{10!}\times\tfrac{1}{80!}$).

The random permutations considered, and especially the use of the RBS in operation 6, eliminate the patterns sought in cryptanalysis work. Of course, in operation 6 a different RBS must be used systematically and unavoidably for each of the $S_i$.

To be certain about the genuine randomness of each RBS used, one must verify that each sequence passes the most rigorous randomness tests available: \cite{nist} and \cite{irnd}.

In future articles, consideration will be given to the use of several variants of the IRE for secure communication between users who do not share an $S_i$.\\

\end{document}